\documentclass[12pt]{article}

\usepackage[utf8]{inputenc}
\usepackage[english]{babel}
\usepackage[margin=2.3cm]{geometry} %
\usepackage{graphicx} %
\usepackage{amsthm, amsmath, amssymb} 
\usepackage{tcolorbox} 
\usepackage[numbers]{natbib} %
\usepackage{lineno}	
\usepackage{textgreek}
\usepackage{appendix} %
\usepackage{float}
\usepackage{subcaption}

\usepackage{upgreek}\usepackage{xr}
\usepackage{algorithm}
\usepackage{algpseudocode}
\usepackage{authblk} 
 \usepackage{setspace} %
\usepackage{pifont}
\newcommand{\open}{\item[$\square$]}

\usepackage{hyperref}

\newcommand{\papertitle}{Element-deletion-enhanced digital image correlation for automated crack detection and tracking in lattice materials}
\newcommand{\ie}{{\textit{i.e.}}}
\newcommand{\eg}{{\textit{e.g.}}}

\newcommand{\area}{A}

\newcommand{\height}{h}
\newcommand{\volume}{V}
 \graphicspath{ {./} }
\begin{document}

\title{\papertitle}
\author[1]{Alessandra Lingua}
\author[1]{Arturo Chao Correas}
\author[2]{François Hild} %

\author[1]{David~S.~Kammer\thanks{Corresponding Author: dkammer@ethz.ch}}

\affil[1]{Institute for Building Materials, ETH Zurich, Switzerland}
\affil[2]{Université Paris-Saclay, CentraleSupélec, ENS Paris-Saclay, CNRS\\ LMPS--Laboratoire~de~Mécanique~Paris-Saclay, Gif-sur-Yvette, France} 
\maketitle

\section*{Abstract}
Architected materials can exhibit remarkable combinations of stiffness, strength, and toughness, yet their application is currently limited by an incomplete understanding of how cracks initiate and propagate through their discrete architecture. Elucidating the mechanisms that underpin these processes is challenging because lattice failure is governed by highly localized deformations of slender beams, which fall outside the resolution and assumptions of optical methods developed for continuum solids, such as digital image correlation (DIC). Thus, characterizing crack propagation within lattice materials requires measurement strategies capable of resolving lattice-scale deformations while accounting for both the intrinsic discreteness of lattice architectures and the progressive formation of material discontinuities during failure.

This work introduces a global DIC framework tailored to architected materials, in which the correlation problem is solved directly on the lattice mesh and damaged elements are automatically removed during the analyses. Damage detection, which relies on a data-driven residual criterion, enables the robust tracking of localized deformation and crack-tip motion under different testing conditions. The method provides physically consistent displacement field measurements on the evolving intact lattice topology and resolves the crack path over time. Validations on 3D-printed regular and imperfect triangular lattices under mode-I loading demonstrate that the approach accurately captures both damage initiation and crack propagation. Furthermore, we demonstrate that identifying damaged elements provides an estimate of the critical failure strain, which can be used directly in numerical models or adopted as an alternative element-deletion threshold in DIC analyses.

\vspace{2mm}\noindent \textbf{Keywords:}
Additive manufacturing; Architected materials; Damage detection; Digital image correlation (DIC); Fracture characterization 
\section{Introduction} \label{sec:introduction}

Lattice materials find applications across a wide range of fields (\eg, automotive, aerospace, and biomedicine) where the combination of high mechanical performance and low density is essential~\cite{jiao_23}. Engineering their discrete architecture allows one to tailor the global mechanical~\cite{jia_24, manno_2019}, energy-absorption~\cite{YAVAS2022}, and acoustic~\cite{LEE2023} properties, thereby permitting unprecedented functional performance~\cite{surjadi_2019, montemayor_15}. A central obstacle to fully realizing this potential in engineering applications is the ability to predict their resistance to crack growth and elucidate the underlying local failure mechanisms~\cite{shaikeea_2022}. Although fracture prediction is difficult in any material, it is particularly challenging in lattice materials because crack growth arises from the failure of individual struts that are orders of magnitude smaller than the overall structure. Accurately predicting the resulting crack path morphology and its temporal growth, therefore, requires a technique capable of resolving deformations across these length scales~\cite{JOST2021, radi_2023}.

Optical measurement techniques, such as digital image/volume correlation (DIC/DVC)~\cite{besnard_2006, VANDERESSE2018, HOLMES2023} and photoelasticity~\cite{fulco2025}, provide powerful tools for investigating the multiscale failure behavior of materials, allowing the quantification of localized deformations, and deliver full-field measurements with no physical contact~\cite{wang}. DIC estimates displacements by maximizing the correlation between the speckle patterns of a reference and a deformed image~\cite{schreier, Pan_2009}. This procedure can be performed either independently for subsets of the field of view, often referred to as a local approach, or over the entire domain via a global formulation \cite{besnard_2006, hild_2012, WANG_2016}. One advantage of the global approach is that it enforces, by construction, the continuity of the displacement field, which generally leads to more robust and less noisy results. However, because cracks are characterized by discontinuities in the displacement field and classical global DIC techniques assume this field to be continuous, such approaches are not well-suited to the analysis of crack growth.

A few approaches have been developed to deal with displacement discontinuities in continuum materials using global DIC. These methods typically rely on incorporating fracture mechanics concepts or element-deletion criteria into the DIC analysis. For instance, extended DIC enriches the displacement shape functions to explicitly represent cracks~\cite{CRMECA_2007__335_3_131_0}. However, it does not inherently provide a robust mechanism for automatically identifying the location of the discontinuity. Another way to address this problem is to use integrated DIC, in which the unknown displacement field is described using a mechanically admissible model, such as the Williams series expansion. Because this formulation inherently includes the crack-tip singularity, it can represent displacement discontinuities and enables the identification of material parameters as well as the tracking of the crack tip position over time~\cite{roux09}.
Although these methods are effective for continuous materials, since they are based on standard fracture mechanics, they are not well suited to architected materials with non-negligible size effects, as in this case, the underlying assumptions may not be valid~\cite{SHAIKEEA2024}. Hence, currently existing fracture-mechanics-based DIC methods cannot be used to characterize the failure of lattice materials. An alternative strategy to analyze the displacement discontinuities associated with crack propagation in continuum materials is to integrate an element-deletion criterion within the DIC framework~\cite{CHEN2023}. In this approach, the elements where damage is detected are removed from the mesh, such that the displacement fields are computed solely over the remaining intact material. However, defining physically meaningful and automated criteria for element deletion that remain accurate as the crack propagates is nontrivial, particularly in lattice structures composed of slender beams~\cite{radi_2023, Isola19cmat2, HASSAN20211}. As a result, such an approach has not yet been established for lattice materials.

In this work, we propose a DIC approach with an integrated element-removal criterion that enables tracking of the crack-tip position over time, determination of the crack morphology, and full displacement field measurements over the intact subdomain of the specimen.
Failure events and their spatial locations are inferred from image-wise fluctuations in the DIC residuals.
This data-driven approach requires two DIC analyses, namely, the first analysis determines a robust damage threshold, and the second progressively removes damaged elements.

The identification of damaged elements via the data-driven approach permits the estimation of their critical failure strain. We then assess the applicability of these critical deformations as thresholds in an alternative mechanics-based element-deletion method.
In this approach, elements are removed once their strain exceeds the prescribed threshold. This damage-detection method is more physically grounded and potentially more robust under complex deformation, and requires only a single DIC analysis when the failure properties of the lattice, defined over the same mesh, are known.

A key innovation of our framework is the reliable and fully automated detection of local failures across a wide range of deformation regimes and imaging conditions, including at lattice joints, which conventional methods often fail to resolve.
We validate our DIC-based methodology by considering both regular and imperfect triangular lattice specimens under mode I loading. We find that damaged struts are removed consistently with post-test inspections and element deletion results in significant reductions in the measured residuals, indicating the effectiveness of our proposed methodology. Furthermore, we show that the crack propagation events identified by DIC align with the force drops recorded during the experiment.
In this paper, we first present the experimental setup used in this work and review the fundamental principles of DIC, including the mesh used for correlation analyses. We then describe a methodology for incorporating element deletion within the DIC framework, which extends the possible applications of DIC to crack propagation within lattice-based materials, and present the obtained results. 
Based on strain measurements obtained via DIC with residual-based element deletion, we introduce a mechanics-based element-deletion criterion and assess its accuracy and potential applications.
Finally, we assess the effectiveness of the proposed element-deletion approach by comparing the onset of failure detection and the crack-tip tracking with experimental measurements.

\section{Experimental setup and procedure} \label{sec:exp_methods}

\begin{figure}[t!]
    \centering
    \includegraphics[width=1\linewidth]{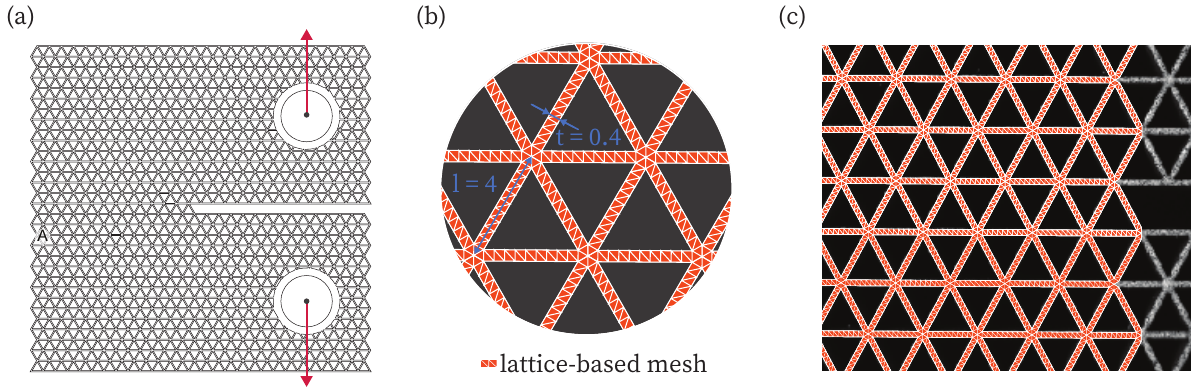}
    \caption{Compact tension specimen design with triangular equilateral cell topology and lattice mesh used in the DIC analyses. (a)~The specimen is subjected to mode I loading via rigid pins inserted in the reinforced holes. (b)~The mesh (orange and white) discretizes each beam into triangular elements and includes triangulated hexagonal joints centered at the lattice network joints. The unit cell characteristic length ($l$) and the lattice beam width ($t$) are highlighted. (c)~Backtracked lattice mesh superimposed on the specimen surface in the vicinity of the notch root.}
    \label{fig:1}
\end{figure}

To validate the DIC approach proposed for architected materials, we manufactured and tested compact tension specimens with a triangular lattice topology. We 3D-printed the specimens with a geometry adapted from the ASTM E-1820 standard~\cite{ASTM_E1820}, choosing the stretch-dominated triangular cell topology. The lattice consisted of equilateral triangular cells with a characteristic length of 4~mm and a strut thickness of $\sim0.4$~mm, ensuring a sufficient surface area for reliable pattern matching (Figure~\ref{fig:1}(a)). 

We manufactured the specimens with a Form3 machine using gray V4 resin, both supplied by Formlabs. We washed the specimens with Form wash using isopropanol to remove excess resin and cured them in the Form cure machine. We patterned the specimens using Dupli-Color matte white and black spray paints for correlation purposes. Additional details concerning the manufacturing, curing, and storage of the specimens are reported in Ref.~\cite{lingua2025}. For all the experiments presented in this work, we applied an opening load with an Instron 100~kN testing machine at 2~mm/min stroke-controlled rate (in the direction indicated with the red arrows in Figure~\ref{fig:1}(a)).

\section{Digital image correlation with no element deletion} \label{sec:dic_basics}

We acquired images of the specimen surfaces at a rate of 2~fps using a high-resolution LIMESS camera (Dantec Dynamics) positioned in front of the specimen. We then analyzed the images using the finite-element-based DIC code Correli~3.0~\cite{lecl15}. The displacement fields are estimated by minimizing the gray level difference between a reference ($f$) and a deformed ($g$) image, based on the optical flow equation~\cite{hild_2006}
\begin{equation}
    g(\textbf{x}+\textbf{u}(\textbf{x})) = f(\textbf{x}) + b(\textbf{x}),
\end{equation}
where $b$ is the acquisition noise, and \textbf{u} the unknown displacement at (pixel) location \textbf{x}. 
In this work, a direct correlation approach is adopted to minimize the measurement uncertainty and limit error accumulation, with the first image taken as the unique reference ($f$). This approach is appropriate since brittle lattices do not undergo large deformations prior to failure. The correlation residuals %
\begin{equation}
    \uprho_{\textbf{u}}(\textbf{x}) = f(\textbf{x}) - g(\textbf{x} + \textbf{u}(\textbf{x})) ~
\end{equation}
provide an estimate of the image matching success. High correlation residuals indicate the occurrence of material failure~\cite{hild15}.
In global DIC, the minimization of the sum of squared gray-level residuals is performed over the entire meshed surface at each time step, ensuring displacement-field continuity and facilitating the comparison with finite-element simulations that can often be undertaken using the same mesh~\cite{Leclerc09,Rannou10}. 

To localize and track damage over a discrete network, we performed the DIC analysis on a lattice-based mesh composed of beams discretized into eighteen uniformly sized triangular elements and connected through hexagonal joints, where each joint is meshed using six triangles (Figure~\ref{fig:1}(b,c)). This approach ensures a constant area for each mesh element across both the beams and joints, so that element-wise measurements can be directly compared. As a result, damage can be localized at lattice joints or along beams in an unbiased manner based on gray-scale or residual variations. In the present work, the gray-scale intensity ranges from 0 to 255.
The mesh is backtracked over the acquired reference image to ensure accurate superposition with the lattice geometry and precise node tracking during mechanical testing (see Appendix~\ref{appendix:backtrack} for details on the backtracking procedure). %
The displacement fields obtained via DIC are used to compute the deformation gradients within each triangular element using linear shape functions. From these gradients, the Green–Lagrange strain tensor is evaluated, and its maximum principal values ($\upvarepsilon_1$) are extracted.

\begin{figure}[!t]
    \centering
    \includegraphics[width=1\linewidth]{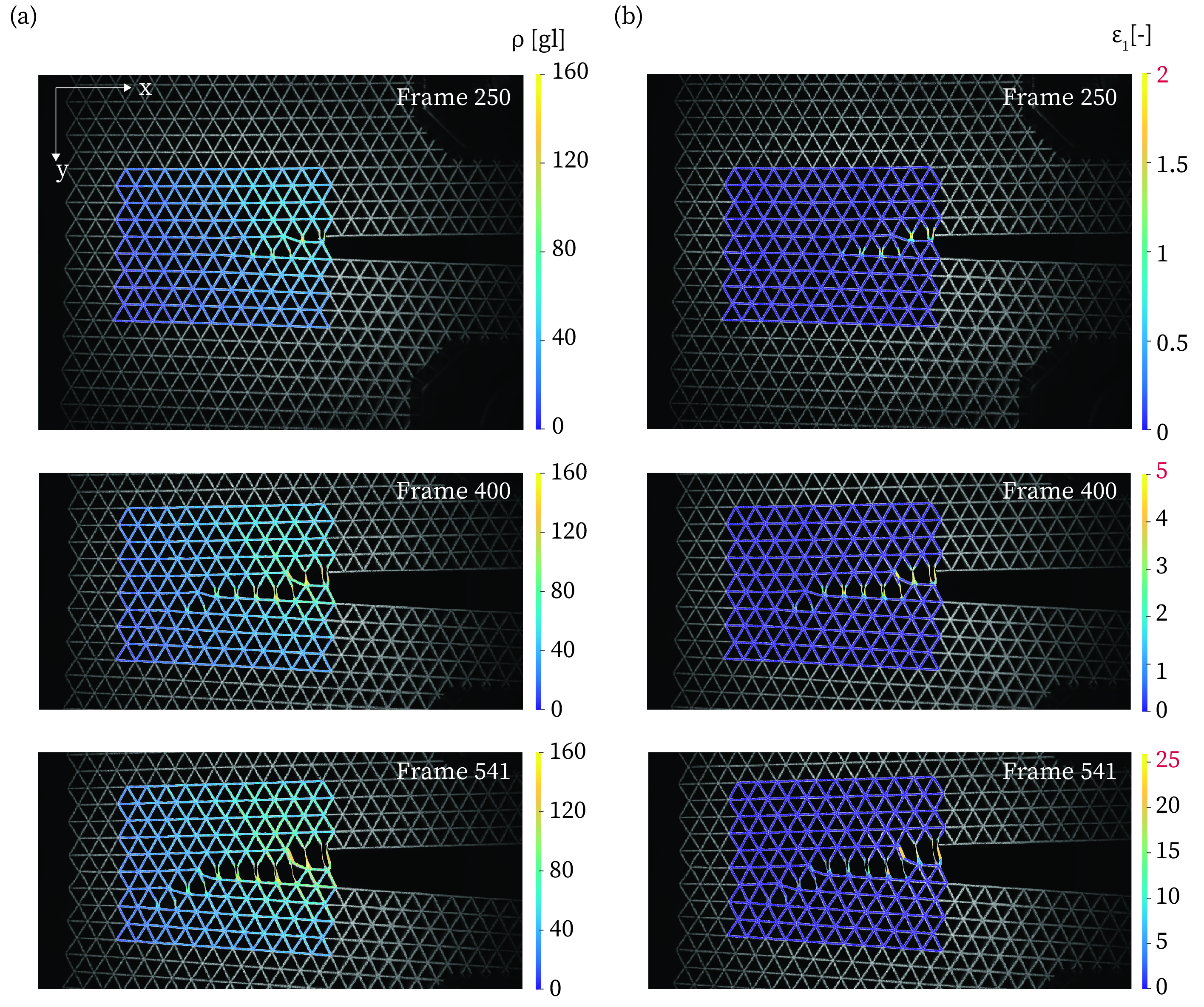}
    \caption{Full-field measurements over the lattice mesh during crack propagation up to the end of test (frame 541) obtained by global DIC with no element deletion. (a) Residual ($\uprho$) field. (b) Maximum principal strain ($\upvarepsilon_1$) field. High residuals and unphysical strains are observed along the propagation path in the vicinity of the notch, as no damage criterion is applied. The scale bar has to be adjusted due to the progressive increase in deformations.}
    \label{fig:3}
\end{figure}

The DIC analysis performed with \emph{no element deletion} provides continuous residual fields over the lattice mesh (Figure~\ref{fig:3}(a)). Because no damage mechanism is incorporated in the kinematic description, the elements across the crack path are assumed to continue to deform as the lattice fractures. This leads to the calculation of excessive, physically unrealistic strains $\upvarepsilon_1$ (Figure~\ref{fig:3}(b)).
While the final crack path and failure pattern can be inferred a posteriori by identifying highly deformed elements in the last acquired image, this approach does not allow for the reliable tracking of the crack tip throughout the loading history. In principle, damage could be inferred from the residual field, but no robust method exists to define a threshold that works consistently over the entire test. Defining strain-based criteria for damage detection is even more complex, as deformations progressively increase with crack opening. It is thus challenging to set a fixed threshold that systematically identifies damaged elements unless they are progressively removed within the DIC framework.

\section{Methods: Element-deletion approach} \label{sec:dic_methods}

We propose a data-driven element-deletion approach in which residual variations exceeding a prescribed threshold are interpreted as being indicators of damage. The method consists of two passes. In the \emph{first pass}, a standard DIC analysis is performed to determine an appropriate residual-variation threshold for the detection of both damage events and failure loci. In the \emph{second pass}, elements are removed based on their residual variations computed image-wise. The crack topology and crack tip location are tracked by extracting the $x$ and $y$ coordinates of the centroids of the removed elements.

\subsection{First pass: DIC with no element deletion to identify the residual threshold}\label{sec:first_pass}

We first perform a classical DIC analysis with no element deletion to determine a residual-variation threshold suitable for damage detection. We then analyze the evolution of the residual field across the image sequence. 

For the $i-\mathrm{th}$ deformed image, we compute the incremental residual variation at the element level, which is given by
\begin{equation}
  \Delta\uprho_i^e = \uprho_i^e - \uprho_{i-1}^e ~,  
\end{equation}
where $e$ denotes the element. For each image, we then retain the maximum variation over all elements,
\begin{equation}
   \Delta\uprho_i^\mathrm{max} = \max_e \left(\Delta \uprho_i^e\right) ~. 
\end{equation}

\begin{figure}[t!]
    \centering  
    \includegraphics[width=1\linewidth]{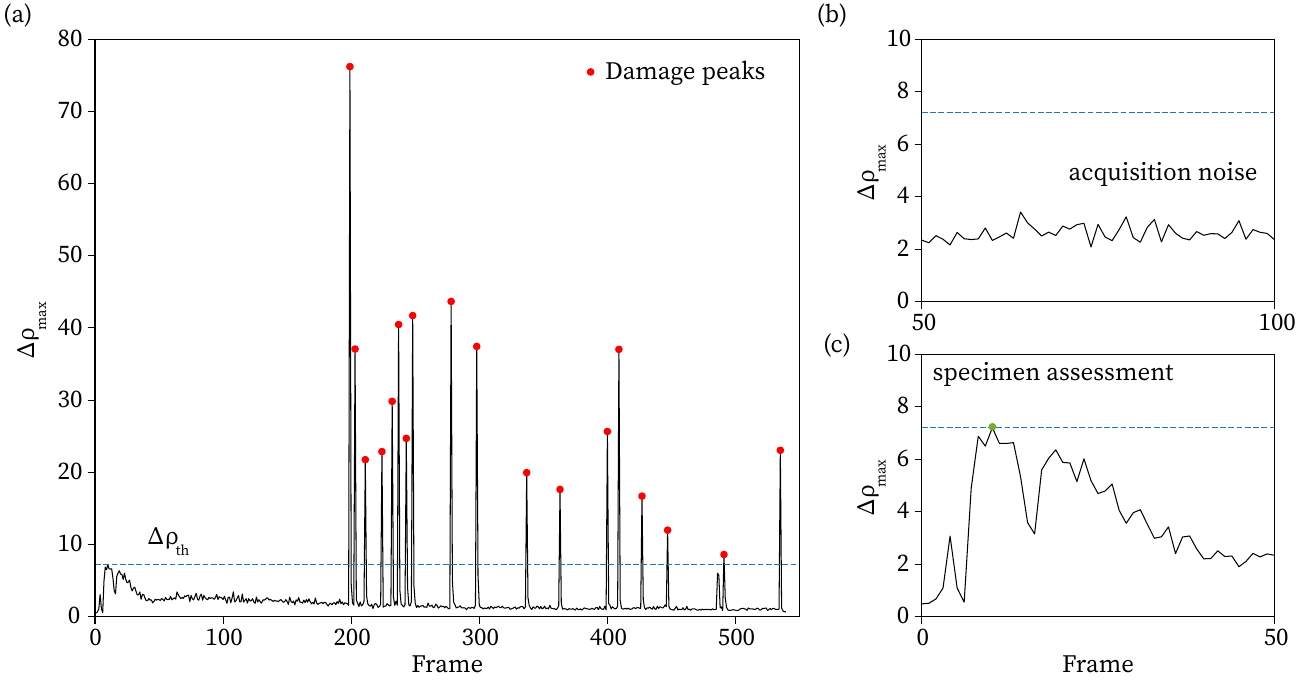}
     \caption{Analysis of the maximum residual variation to identify a threshold that distinguishes damage events (isolated peaks) from acquisition noise. (a)~Maximum incremental residual variation $\Delta\uprho_i^\mathrm{max}$ as a function of frame index $i$. The isolated peaks corresponding to damage events identified by the method presented here are marked with red dots. (b)~The peak detection method identifies and ignores the background noise based on signal prominence. (c)~The minimum relative distance between neighboring peaks ensures that the residual variations originating from the initial specimen adjustment in the grips are not interpreted as damage. The highest peak not associated with a damage event is used to define the threshold $\Delta\uprho_\mathrm{th}$ = 7.3 gl, obtained by increasing this peak by 1\% (blue dashed line).}
    \label{fig:2}
\end{figure}

The signal $\Delta\uprho^\mathrm{max}_{i}$ exhibits a well-defined baseline of around 2 gray levels (gl), %
interrupted by highly localized peaks, see Figure~\ref{fig:2}(a). This baseline reflects the intrinsic noise associated with image acquisition, and is taken to represent the level of residual fluctuations in the absence of damage. By contrast, the pronounced peaks indicate abrupt increases in the residual in at least one element, which correspond to discrete damage events.

To separate damage-related peaks from fluctuations caused by noise (Figure~\ref{fig:2}(b)) or global brightness variations (Figure~\ref{fig:2}(c)), we estimate the noise level in $\Delta\uprho_i^\mathrm{max}$ using the following procedure. We first compute the median absolute deviation (MAD) of the signal $\Delta\uprho_i^\mathrm{max}$,
\begin{equation}
    \mathrm{MAD} = 
\mathrm{median}\!\left( 
\left| \Delta \uprho_i^{\max} - \mathrm{median}(\Delta \uprho_i^{\max}) \right| 
\right) ~,
\label{eq:mad}
\end{equation}
and define the associated noise scale, assuming a Gaussian distribution, as 
\begin{equation}
\sigma_\mathrm{noise} = 1.4826 \cdot\mathrm{MAD}.
\label{eq:sigma_noise}
\end{equation}
We then use this value of $\sigma_\mathrm{noise}$ to identify statistically significant peaks in the $\Delta\uprho_i^\mathrm{max}$ signal. A peak at image index $i$ is classified as a damage event if two conditions are satisfied. First, its prominence relative to the surrounding local minima exceeds $4\sigma_\mathrm{noise}$. Second, its vertical distance to the immediately adjacent data points exceeds $2\sigma_\mathrm{noise}$. These criteria ignore gradual or broadly distributed signal variations, such as those arising from specimen settling in the grips or slow grayscale drift, and identify only abrupt residual increases that are clearly distinct from the background noise.

We denote by $\mathcal{P}$ the set of indices satisfying these conditions. The threshold $\Delta\uprho_\mathrm{th}$ is then defined as the maximum value of $\Delta\uprho_i^\mathrm{max}$ after excluding the indices in $\mathcal{P}$, increased by $1\%$. This value represents an upper bound of residual fluctuations that are not attributed to damage. For the tested triangular lattice specimen, the analysis of the maximum image-wise residual variation leads to an element deletion threshold of $\Delta\uprho_\mathrm{th} = 7.3$~gl. This threshold successfully captures all damage events (see Fig.~\ref{fig:2}(a)), with the exception of a single peak around frame 480 that falls below $\Delta\uprho_\mathrm{th}$. Neglecting this minor event is unavoidable in order to exclude residual variations associated with the initial specimen assessment, and does not compromise the tracking of crack propagation.

\begin{comment}
\subsubsection{Results} \label{sec:rho_results}

The DIC analysis performed with \emph{no element deletion} provides a continuous displacement field over the entire mesh together with the associated residual field $\uprho$. 
%
%
Displacement field continuity is assumed by using a global DIC approach, while mesh backtracking guarantees that elements deform consistently with the lattice. 
From the displacement field, we compute the maximum principal strain field $\epsilon_1$. 

%
%

At the end of the test, both the residual and the principal strain fields exhibit pronounced localized maxima along the crack path and in the vicinity of the notch, see Figure~\ref{fig:3}. These regions coincide with the zones where damage and fracture develop.

Because no damage mechanism is incorporated in the kinematic description, the elements spanning the notch exhibit high residuals (Figure~\ref{fig:3}(a)) and continue to deform as the crack opens up to excessive, physically unrealistic strains (Figure~\ref{fig:3}(b)).

%
While the end-of-test crack path and final failure could be inferred by identifying the highly deformed elements in the last acquired image, this approach does not provide the crack tip position at each time step.
Following the procedure detailed in Section~\ref{sec:rho_method}, we analyze the maximum elementary residual variation image-wise. By identifying the maximum non-isolated peak and by increasing it by 1\% of its value, we obtain a threshold $\Delta\uprho_{th}$ = 7.3.
\end{comment}
\subsection{Second pass: DIC with data-based element deletion}
\label{sec:second pass}
\label{second_pass}
After having obtained the residual variation threshold ($\Delta\uprho_{\mathrm{th}}$), we perform a \emph{second DIC analysis} in which damaged elements are removed if their residual variation at time $t$ exceeds $\Delta\uprho_{\mathrm{th}}$ \ie:
\begin{equation}
   \mathcal{D}(t) = \{ i \mid \Delta \uprho_i(t) > \Delta \uprho_{\mathrm{th}} \},
   \label{residual_damage_criterion}
\end{equation}
where $\mathcal{D}(t)$ is the set containing all damaged elements at time $t$. At each time step where damage is detected, the mesh is modified by removing the elements belonging to the damaged elements set $\mathcal{D}(t)$ and the connectivity matrix is updated accordingly. The labels of the damaged elements are stored for later postprocessing. 

\begin{figure}[!b]
    \centering
    \includegraphics[width=1\linewidth]{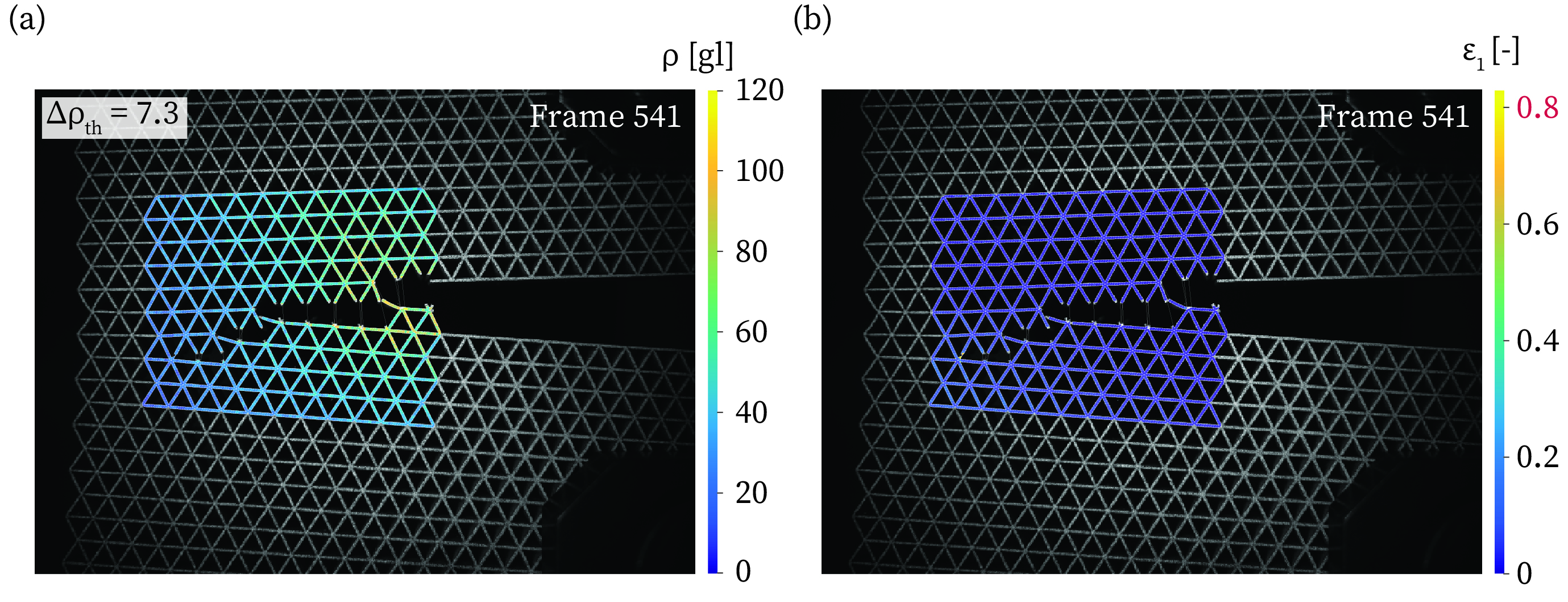}
    \caption{End-of-test $\uprho$ and $\upvarepsilon_1$ fields for the lattice-based mesh with integrated element deletion based on $\Delta\uprho_{th} = 7.3$ gl. The damaged element deletion yields a significant reduction in the residuals~(a) and local deformations~(b) maxima compared with the values obtained via an analysis undertaken without element deletion (Figure~\ref{fig:3})}
    \label{fig:temp}
\end{figure}

As shown in Figure~\ref{fig:3}, standard global DIC yielded unphysical measurements across the notch where the crack propagated. Element deletion based on $\Delta\uprho_{th}$ allows the correlation problem to be solved exclusively where the material is intact and thus leads to a reduction of the residual and strain maxima in the full fields, as shown in Figure~\ref{fig:temp}.
In particular, maximum elemental strains ($\upvarepsilon_1 \sim 0.8$ at the end of test) are around one and a half orders of magnitude lower than those calculated with no element deletion (Figure~\ref{fig:3}(b)),  highlighting the effectiveness of progressive element deletion based on an image-wise residual variation threshold.

Additionally, our method allows us to extract the elemental strain at failure for each event from the previous time step ($t-1$). 
By mapping the mesh elements to the lattice strut they belong to, we estimate the average failure strain per strut ($\upvarepsilon_{crit}$) and the number of struts failing at each time step.
\begin{figure}[!b]
    \centering
    \includegraphics[width=1\linewidth]{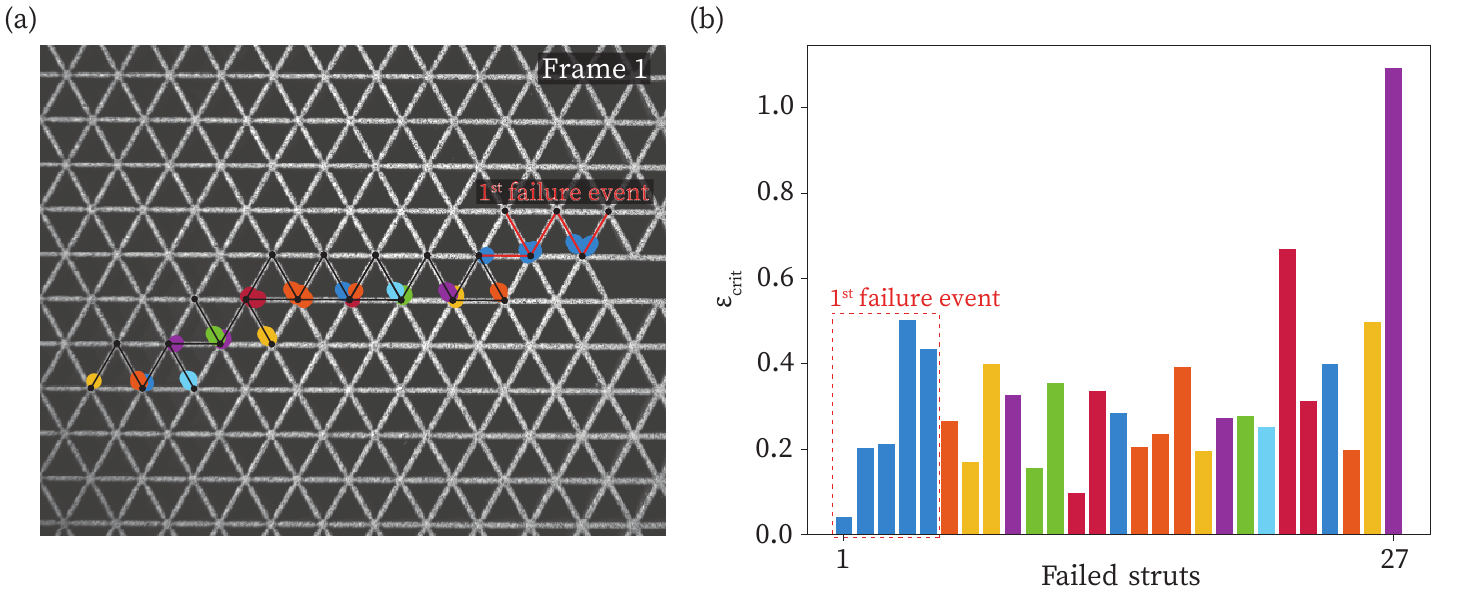}
    \caption{Damaged elements identified using the residual-based criterion and the average critical strain per strut. (a)~Damaged elements detected with the threshold $\Delta\uprho_{th}$ = 7.3~gl are shown over the lattice in the undeformed configuration. (b)~Mapping damaged elements to their respective struts allows the average critical strain per strut to be estimated, as well as the number of damaged struts during the test. The color in (a) and (b) refers to the sequence of the failure events.}
    \label{fig:4}
\end{figure}

\sloppy For the tested triangular specimen, the DIC analysis with element deletion based on ${\Delta\uprho_{th} = 7.3~\mathrm{gl}}$ leads to the identification of the damaged elements shown in Figure~\ref{fig:4}(a). We observe that failure occurs in the vicinity of the lattice joints, which is likely due to stress concentrations at sharp corners. 
Mapping the elements to the lattice strut they belong to allows us to estimate directly from DIC measurements the average critical strain per strut and the global number of failed struts in the specimen, which, at the end of the test, is estimated to be 27 (Figure~\ref{fig:4}(b)). 
\begin{figure}[!b]
    \centering
    \includegraphics[width=1\linewidth]{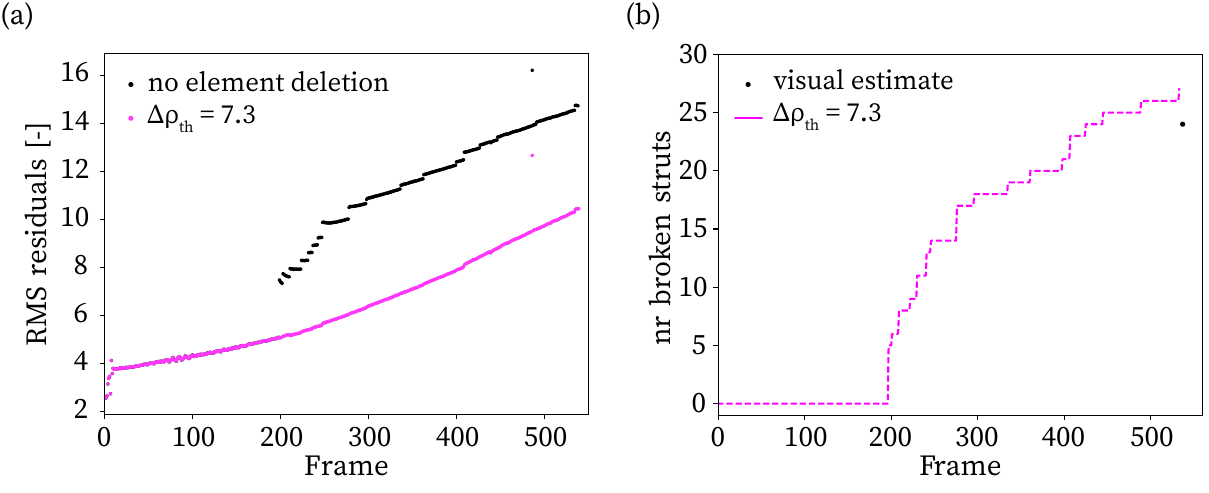}
    \caption{Evolution of global DIC residuals with and without element deletion and estimated number of broken struts during the test as a function of frame number. (a)~Global RMS residuals increase similarly for both approaches until the onset of damage at frame 200, after which they rise sharply when no element deletion is applied. (b)~Cumulative number of broken struts plotted versus image number, up to an end-of-test estimate of 27. The number of broken struts identified by visual inspection at the end of the test is reported for comparison purposes (24 struts).}
    \label{fig:5}
\end{figure}

The performance of the proposed framework is assessed via root mean square (RMS) residuals, which quantify the convergence error of the DIC solver as the loading evolves. The RMS residuals increase steadily in both the standard DIC analysis and in the analysis with element deletion up to the onset of failure, at image 200 (see Fig.~\ref{fig:5}(a)). Thereafter, the residuals rise sharply due to damage initiation when no element-deletion strategy is applied. Progressive element deletion based on the threshold $\Delta\uprho_{th}$ leads to a pronounced reduction in the global residuals compared with the classical DIC analysis (Figure~\ref{fig:5}(a)), indicating improved correlation fidelity during damage evolution.
Furthermore, removing damaged elements improves measurement accuracy by limiting the gray-scale range within the images to be correlated (Figure~\ref{fig:temp}(a)) compared to the analysis with no element removal (Figure~\ref{fig:3}(a)). The number of broken struts detected by the DIC analysis with element deletion increases from frame 200 onward up to 27 broken struts at the end of test (Figure~\ref{fig:5}(b)). Visual inspection at the end of the test reveals 24 broken struts, indicating the accuracy of the automated DIC-based element-removal strategy. The slight discrepancy arises because each joint is considered as being shared among multiple struts; when damage is detected at one joint, all struts connected to it are counted as broken during post-processing, including those aligned with the $x$ direction.

\section{Approach validation: crack path detection in irregular lattices} 
\label{sec:imperfection}
\begin{figure}[!b]
    \centering
    \includegraphics[width=.9\linewidth]{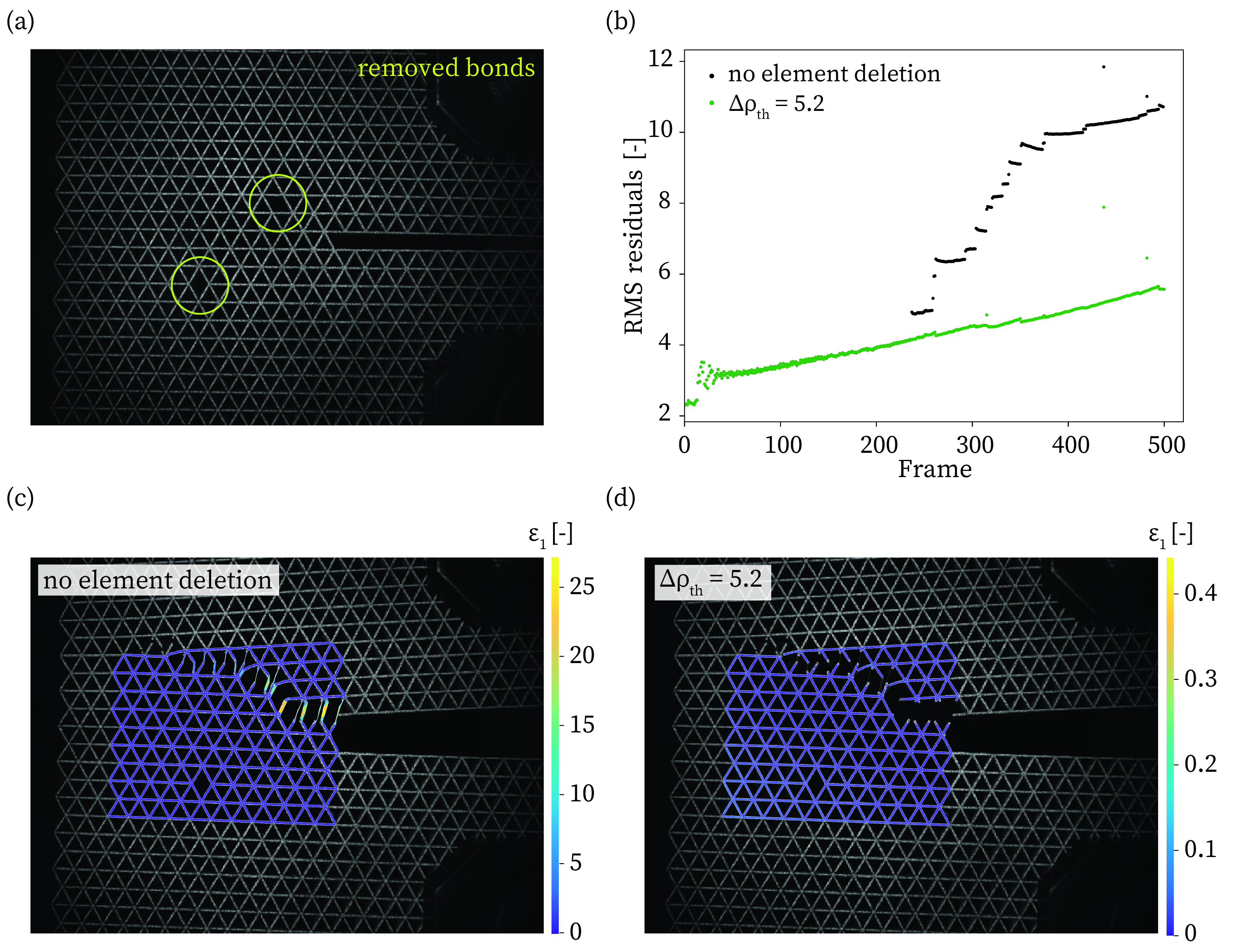}
    \caption{Strain field at the end of test for a specimen with imperfections in the form of missing struts, and global residual changes with and without element deletion. (a)~The location of the removed struts is highlighted with yellow circles. (b)~Element deletion based on residual variations leads to a reduction in global residuals after the onset of failure (frame 238). (c)~The elements along the notch deform to an unphysical extent when no deletion is applied. (d)~The automatic detection and deletion of broken elements results in DIC being performed exclusively on the intact mesh and yields strain levels up to an order of magnitude lower than those obtained with classical DIC.}
    \label{fig:11}
\end{figure}
To further validate the proposed element-deletion approach, we assess its performance in tracking complex, tortuous crack paths in an imperfect lattice specimen under mode I loading. The imperfection is introduced by randomly removing two struts within the triangular lattice tesselation (Figure~\ref{fig:11}(a)). Following the procedure detailed in Section~\ref{sec:first_pass}, we first perform a classical DIC analysis over a mesh adapted to the imperfect lattice, and we identify the residual threshold for element deletion as $\Delta\uprho_{th}$ = 5.2~gl. To mitigate the higher noise level in the acquired signal, a minimum distance of one frame is enforced between a detected peak and the maximum non-peak value, thereby limiting the influence of noise near damage-related peaks.

We then perform a second DIC analysis following the methodology presented in Section~\ref{sec:second pass} with element deletion occurring when the local residuals exceed $\Delta\uprho_{th}$. By post-processing the failure strain of the element detected as broken, we estimate the average critical failure strain as $\upvarepsilon_{crit}$ = avg$(\upvarepsilon_{avg})$ = 0.28 $\pm 0.1$ (standard deviation) in the presence of imperfections.

The residual and strain fields at the end of test revealed that the interaction between the propagating crack and the vacancy introduced by design resulted in the deviation from the notch plane (Figure~\ref{fig:11}(c,d)). The proposed method successfully detected damage even along a more complex path, and successfully tracked the crack tip position evolution in time along both the $x$ and $y$ direction (Figure~\ref{fig:11}(d)).
\section{Discussion} \label{sec:discussion}
The proposed approach for integrating damage detection into the DIC framework successfully removes broken elements from the mesh at each time step, enabling the continuous tracking of crack-tip motion and generating physically meaningful displacement and strain fields over the intact lattice. In addition, introducing a damage criterion makes it possible to quantify the number of struts that fail throughout the test and to follow the crack-tip evolution, $da$, in time, even when the crack path is complex (the quantity $da$ is approximated here by the $x$-coordinate of the most advanced damaged element).
Ultimately, the identification of the damaged elements based on their residual variation during testing allows us to derive an estimate of the elemental failure strain. The availability of an empirical estimate for the failure strain suggests the possibility of identifying damaged elements based on their elemental deformations, as an alternative approach to the residual-based method described above. 
We note that all measured critical strains represent a lower bound for the strain at failure due to the limitation of image acquisition rate (two images per second), which may not capture the true strain at the moment of failure.
This mechanics-based approach may be particularly suitable when the experimental conditions introduce significant grayscale variations that preclude the definition of a robust residual-based damage threshold.
The initial estimate of the strain threshold obtained from the data-driven procedure can also be refined to ensure an accurate representation of crack propagation. Alternatively, it can be directly replaced by the failure strain evaluated at the lattice mesh element level, if known, thereby reducing the number of required DIC passes from two to one.

In the following, we describe how the mechanics-based thresholds can be obtained from the residual-based DIC analysis, and assess the degree to which the two methods agree.
\subsection{Alternative mechanics-based element deletion DIC approach}
To infer a mechanics-based damage threshold from the failure strains of elements as measured using the residual-based damage-identification criterion (\ie,~Eq.~(\ref{residual_damage_criterion})), we evaluate both the maximum and the average critical elemental deformation for each failed strut. Considering these strain values in a failure-event-wise manner (including failure events in which multiple struts fail), we then extract three distinct metrics: the average of the maximum (avg($\upvarepsilon_{\mathrm{max}}$)) and average strain (avg($\upvarepsilon_{\mathrm{avg}}$)), and the maximum of the maximum strain (max($\upvarepsilon_{\mathrm{max}}$)) strut-wise.
We take these three values as strain thresholds ($\upvarepsilon_{\mathrm{th}}$) for a mechanics-based element removal process.
We then run a new element-removal-enhanced DIC analysis, during which the elements whose strain exceeds $\upvarepsilon_{th}$ are added to the set $\mathcal{D}(t)$ and are irreversibly deleted from the mesh:
\begin{equation}
   \mathcal{D}(t) = \{ i \mid \Delta \upvarepsilon_i(t) > \upvarepsilon_{\mathrm{th}} \}~.
\end{equation}
The connectivity matrix is updated accordingly at each time step in which damage is detected, and the indices of the deleted elements are stored for subsequent analyses.

\begin{figure}[!t]
    \centering    \includegraphics[width=1\linewidth]{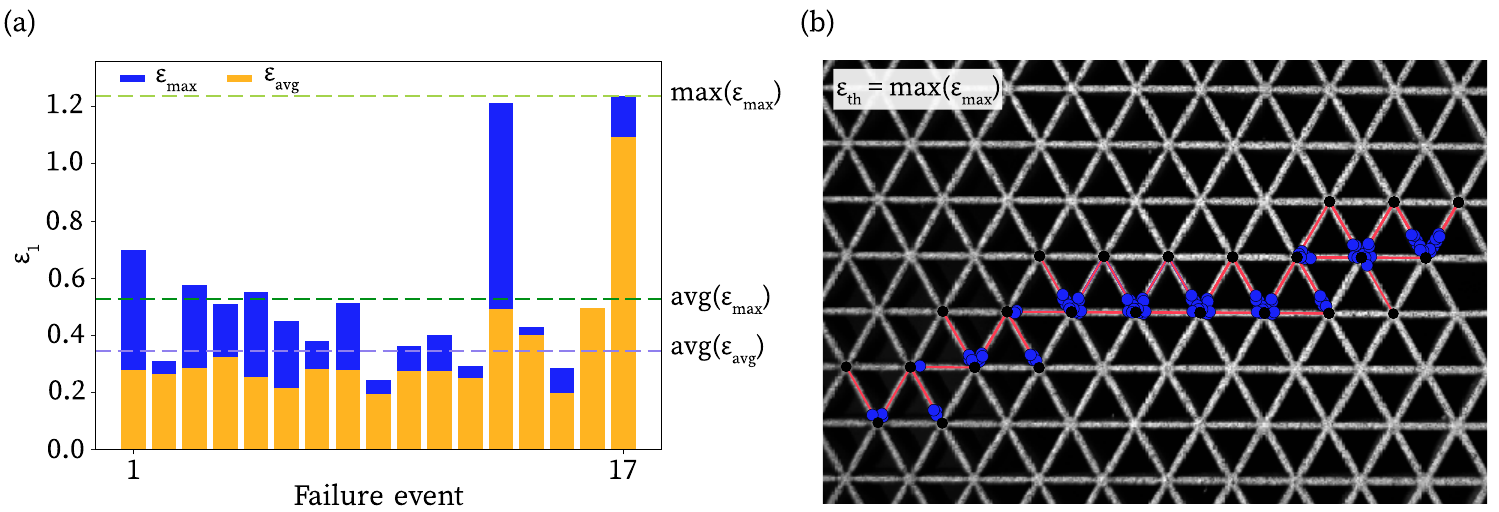}
    \caption{Strain-threshold identification from the critical deformations of elements removed by the residual criterion, and corresponding crack path found through DIC with strain-based element deletion. (a)~The average and maximum critical strain is computed for each failed strut, and their event-wise mean and peak value provides estimates of the failure strain for each of the seventeen failure events. We report in orange the event-wise average of the average strain per strut, and in blue the maximum of the peak values per strut, event wise. The average of the average strain at failure is 0.34 (dashed violet line), while the average and maximum of the peak values yield the strain-based thresholds $\epsilon_{th} = 0.5$ (dark green dashed line) and 1.2 (light green dashed line), respectively. (b)~Applying $\epsilon_{th} = 1.2$ as threshold in a subsequent DIC analysis with strain-based element deletion results in the failure path shown, with damaged elements identified in the vicinity of the nodes.} %
    \label{fig:6}
\end{figure}

For the triangular lattice analyzed here as a representative example, a residual threshold of $\Delta\uprho_\mathrm{th} = 7.3$~gl yields the critical strain values for all broken struts illustrated in Figure~\ref{fig:4}{b}. By computing the average and maximum strain in a strut-wise manner, and for each failure event, we obtain avg($\upvarepsilon_{\mathrm{avg}}$) = 0.34, avg($\upvarepsilon_{\mathrm{max}}$) = 0.5, and max($\upvarepsilon_{\mathrm{max}}$) = 1.2 (see Figure~\ref{fig:6}(a)).
To highlight the possibility of using a mechanics-based element-removal procedure, we performed the DIC analysis with strain-based element removal with the three thresholds: avg($\upvarepsilon_{\mathrm{avg}}$), avg($\upvarepsilon_{\text{max}}$), and max($\upvarepsilon_{\text{max}}$).

\begin{figure}[!h]
    \centering
    \includegraphics[width=1\linewidth]{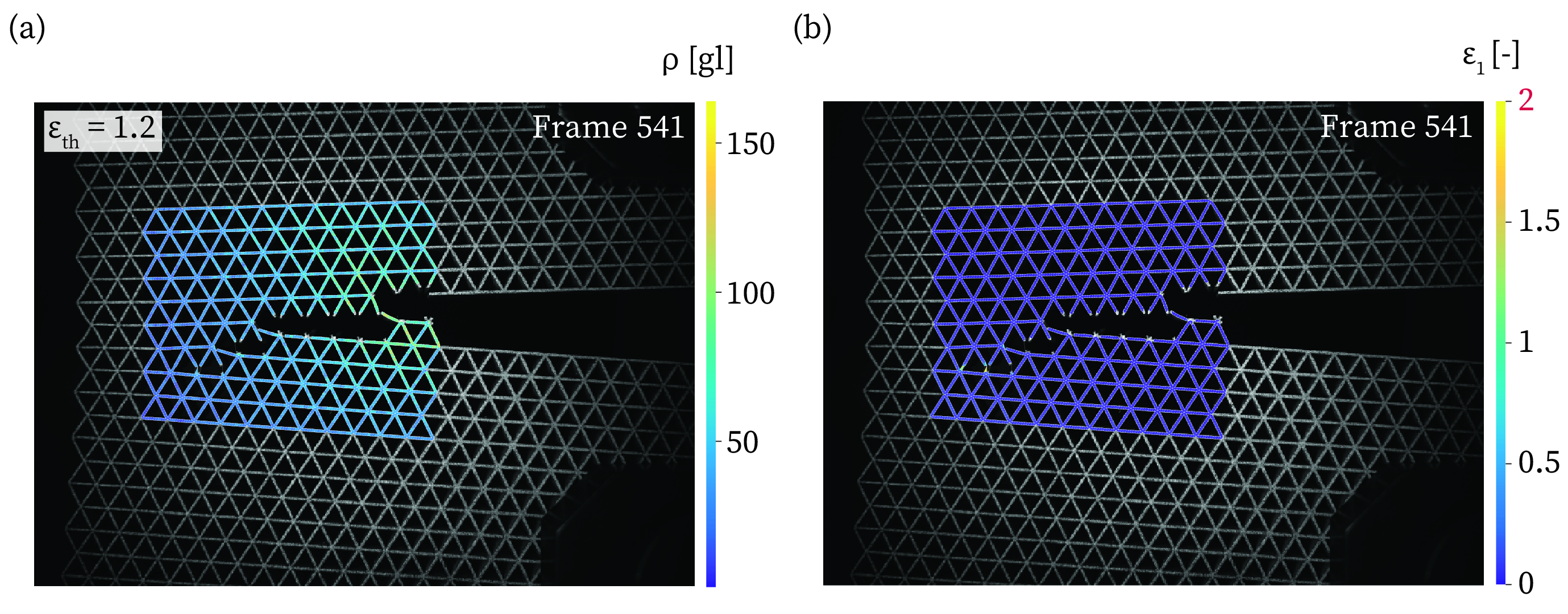}
    \caption{End-of-test $\uprho$ and $\upvarepsilon_1$ fields by DIC with integrated element deletion based on the threshold $\upvarepsilon_{th} = 1.2$. The damaged element deletion yields a reduction in the residuals~(a) and local deformations~(b) maxima compared with the analysis without element deletion (Figure~\ref{fig:2}).}
    \label{fig:7}
\end{figure}

Integrating the strain-based criterion into the DIC analysis results in the deletion of elements located near the lattice joints, as shown in Figure~\ref{fig:6}(b) and Figure~\ref{fig:7}, where we take $\upvarepsilon_{th} =  \text{max}(\upvarepsilon_{\mathrm{max}}) = 1.2$ as a representative case.
To assess the influence of the selected strain threshold values, we examine the reduction in RMS residuals resulting from the deletion of damaged elements, as well as the progressive increase in the number of struts identified as being damaged.

\begin{figure}[!h]
    \centering    \includegraphics[width=1\textwidth]{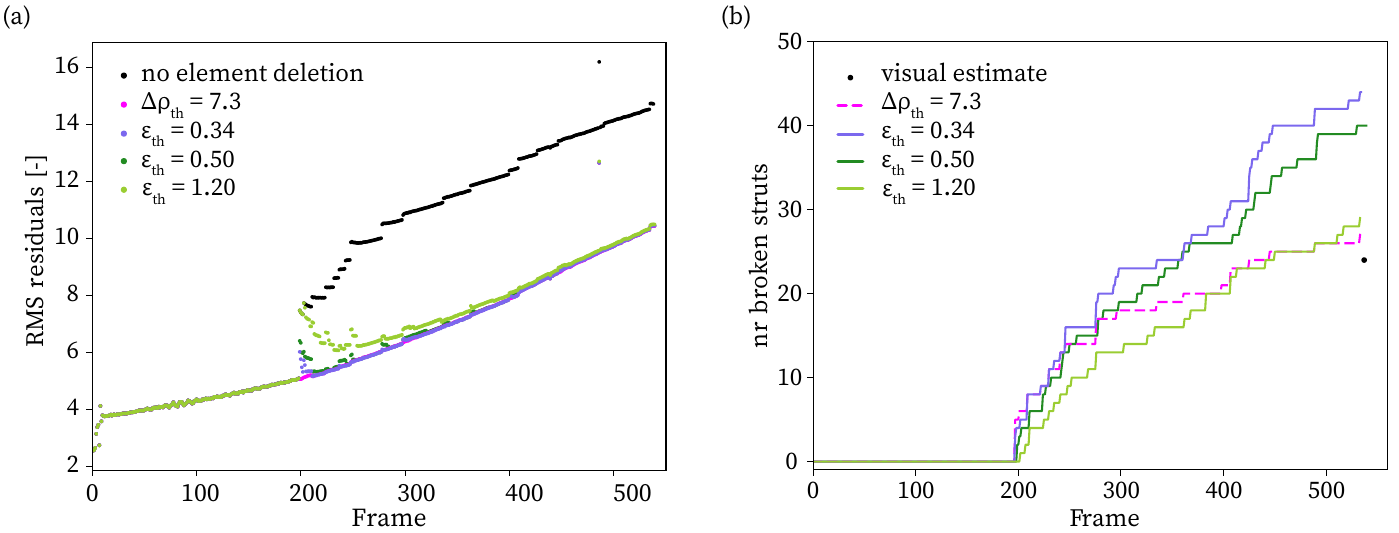}
    \caption{Comparison of RMS residuals and evolution of the number of broken struts found using the data-based and the mechanics-based element deletion methods. (a)~The residual-based criterion yields the most pronounced drop in global RMS residuals, followed by the lowest value of $\upvarepsilon_{th}$ considered here (0.34). Increasing values of the critical strain lead to progressively weaker reductions, particularly around the onset of failure. (b)~Residual-based element deletion provides an estimate of the number of broken struts which slightly exceeds the 24 failures identified by visual inspection in the final image. Using the maximum value of $\upvarepsilon_{th}$ the final estimate of the number of broken struts is higher. The higher predicted counts of broken struts for $\upvarepsilon_{th}=0.34$ and 0.5 result from the less spatially-localized damage detection of the deformation-based failure criteria. }
    \label{fig:8}
\end{figure}  
All strain threshold values considered here detect the onset of damage at frame 200, in agreement with the residual-based method.
This point is marked by the deviation of the RMS residuals obtained with element deletion from those computed without element removal (Figure~\ref{fig:8}(a)).
During crack propagation, the largest reduction in RMS residuals is achieved using the data-driven deletion criterion. The lowest strain threshold leads to a slightly smaller reduction compared with the residual-based method, while increasing strain thresholds result in a less pronounced residuals decrease due to the less stringent deformation criterion for failure. Notably, all methods lead to a similar reduction in the residual values toward the end of the experiment. The analysis of the number of damaged struts indicates that the data-driven method provides the closest estimate when compared with post-mortem visual inspection (Figure~\ref{fig:8}(b)).
Different values of $\upvarepsilon_{th}$ lead to markedly different evolutions in the estimated number of broken struts during testing. Generally, strain-based criteria overestimate the number of broken struts because deformation fields are less localized than anomalies captured in correlation residuals. Low strain thresholds further amplify this effect: large deformations near lattice nodes adjacent to the crack path lead to multiple struts converging at a joint being classified as failed (Figure~\ref{fig:8}(b)).
By contrast, $\upvarepsilon_{th}=1.2$ initially underestimates the number of broken struts compared with the data-driven method, but eventually converges to a final value only slightly higher than the visual estimate (29). It should be noted that, when damage is detected using the strain-based criterion, only the elements whose strain exceeds the threshold are removed, while neighboring elements continue deforming until they also reach this threshold. In contrast, the data-driven approach directly detects failure in the element where the gray-scale intensity changes abruptly, and thus leads to a significantly lower number of broken elements.
This difference reflects the fact that strain-based identification is load-dependent and influenced by mesh continuity, whereas residual-based detection relies on localized, abrupt pixel-level changes to detect failure.

\subsection{Crack tip tracking}
\begin{figure}[H]
    \centering
    \includegraphics[width=.6\linewidth]{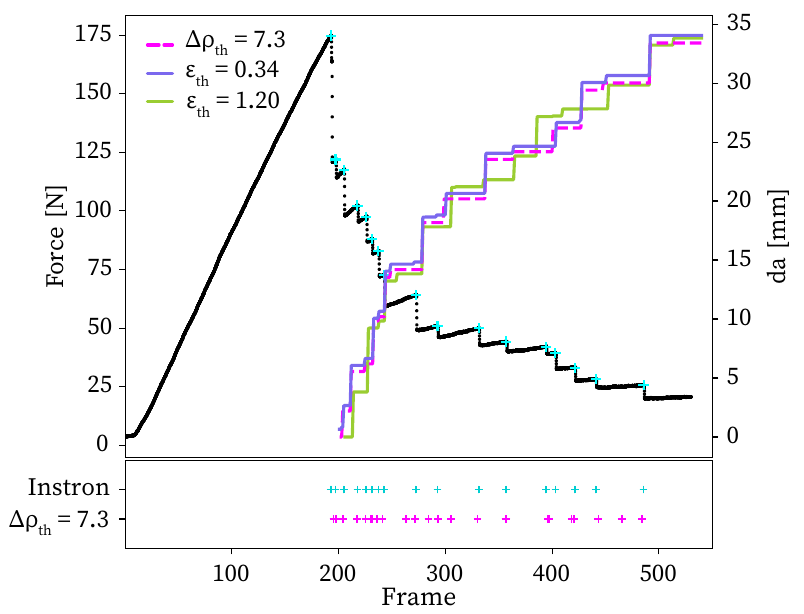}
    \caption{Evolution of the crack-tip position as a function of frame number measured using both element-deletion criteria, and comparison between the timing of damage detection via  DIC and the response recorded by the tensile testing machine. The drops in the recorded force correspond to the damage events detected by DIC.%
    }
    \label{fig:8bis}
\end{figure}

The frame-wise identification of damaged elements allows us to determine the crack tip position as a function of time. Here, we approximate the crack tip position as being the $x$-coordinate of the centroid of the most advanced damaged element.
The crack growth increment, $da$, using both strain- and residual-based thresholding, exhibits consistent behavior (Figure~\ref{fig:8bis}(b) - top). Moreover, the temporal evolution of crack growth measured via the element-deletion-enhanced DIC approach correlates strongly with the drops observed in the force recorded by the testing machine (Figure~\ref{fig:8bis}(b) - bottom), which each correspond to a failure event within the lattice. Minor discrepancies between the image-based detection of damage and the drops in force measurements are likely to be due to differences in measurement frequency and the independent activation of the two data streams at the beginning of the test.

The consistent detection of crack advancement using both data- and mechanics-based element-removal strategies supports the validity of the two approaches. Post-processing the critical strains identified through residual-variation thresholding therefore enables the integration of a physics-based element-deletion criterion within DIC, while also providing an estimate of the elemental critical strain that may be used as a damage parameter in numerical simulations. 
The element-deletion threshold can be selected according to the focus of the analysis; for example, a lower threshold may be adopted when prioritizing crack-path identification over accurate estimation of the number of broken struts
%
%
%
%
%
%
%
\begin{comment}
\begin{figure}[H]
    \centering
        \includegraphics[width=.8\textwidth]{RMS_macro_meso.pdf}
    \caption{Comparison of global residuals for direct analyses with damaged element deletion using strain- and residual-based criteria. (a)~For the cell-based mesh, the residuals progressively increase due to damage if no element is removed (black curve), while introducing a damage criterion significantly reduces the RMS residuals. The strain-based criterion leads to a stronger residual reduction, which corresponds to a better tracking of the crack propagation compared to the residual-based criterion. (b)~For the lattice-based mesh, the same RMS residuals drop is obtained using both criteria for damage.}
    \label{fig:7}
\end{figure}
\end{comment} 
%
%

\section{Conclusion} \label{sec:conclusion}

This study introduced a robust approach for automated damaged-element deletion in global DIC analyses of lattice failure. The methodology not only provided a quantitative estimate of the crack tip position and crack morphology over time in such architected materials, but also physically meaningful displacement and strain fields over the intact specimen at each time step. 

The framework was experimentally validated on 3D-printed regular and imperfect triangular lattices under mode I loading, and was found to successfully capture both the onset and growth of damage, even when the crack path became tortuous due to the interaction with designed imperfections.
Beyond displacement measurements and crack-tip tracking, the proposed framework was shown to provide estimates of the elemental failure strain of the lattice, which can be integrated into numerical models using consistent meshing. We showed that these critical strain values can also serve as an alternative threshold for element deletion within DIC, based on elemental critical deformations, when the failure strain is not known a priori. The possibility of adopting either a residual-based or a strain-based approach represents an additional key feature of the framework, enabling the selection of the most appropriate element-deletion strategy for a given lattice architecture, characteristic scale, and deformation intensity.

Ultimately, the generality of the method facilitates straightforward extensions to disordered architectures, alternative loading modes (\eg mode I–III), and volumetric analyses (\eg, via digital volume correlation), enabling a more comprehensive experimental characterization of fracture in architected materials. 

\section{Acknowledgements}
The authors acknowledge support from the Swiss National Science Foundation under the SNSF Starting Grant (TMSGI2\_211655). The authors thank Dr.~Mohit Pundir and Dr.~Jan Van Dokkum for fruitful discussions and feedback. The authors also acknowledge Dr.~Daniel Rayneau-Kirkhope for feedback on the manuscript draft.
\section{CRediT authorship contribution statement}
\textbf{Alessandra Lingua}: Conceptualization, Methodology, Investigation, Software development, Formal Analysis, Data Curation, Visualization, Writing -- Original Draft. \textbf{Arturo Chao Correas}: Software development, Writing -- Review \& Editing. \textbf{François Hild}: Methodology, Formal Analysis, Software development, Writing -- Review \& Editing. \textbf{David S. Kammer}: Conceptualization, Supervision, Writing -- Review \& Editing, Funding acquisition.

\section{Declaration of competing interest}
The authors declare that they have no known competing financial interests or personal relationships that could have appeared to influence the work reported in this paper.

\section{Data Availability}
The acquired images are available on the ETH research collection: [DOI]

\section*{Appendices}
\appendix
\section{Mesh backtracking} \label{appendix:backtrack}

Figure~\ref{fig:1}(c) shows the lattice-based DIC mesh, which precisely aligns with the lattice topology over the full region of interest for DIC. The mesh was generated following the lattice topology via triangulation and the partitioning of the joints and beams to obtain elements of consistent size. To fit it over the region of interest in the images, a backtracking procedure was followed~\cite{Isola19cmat2}. 

First, the triangular lattice mesh is cropped to match the selected region of interest for DIC analyses (Figure~\ref{fig:3}). From this mesh, a mask was generated in which pixels belonging to the mesh had their gray level set to be equal to the mean gray level of the lattice, while the other pixels had their gray level set to the mean level of the image background. This mask corresponds to the nominal configuration in which the mesh was created. Next, a coarser auxiliary mesh enclosing the region of interest was constructed. An initial DIC analysis was performed to correlate the reference image acquired before loading the specimen and the mask of the nominal configuration. Once convergence was reached, the auxiliary mesh was deformed according to the estimated displacement field and positioned in the nominal frame. Knowing the nodal displacements of the auxiliary mesh, the displacement of any other point from the reference to the nominal configurations was determined via mesh interpolation. The DIC mesh was finally backtracked by applying the opposite of these interpolated displacements to its nodes, thus mapping it from the nominal to the reference configuration. %
\section{Strain threshold sensitivity analysis} 
\label{appendix:threshold}

\begin{figure}[H]
    \centering    \includegraphics[width=1\textwidth]{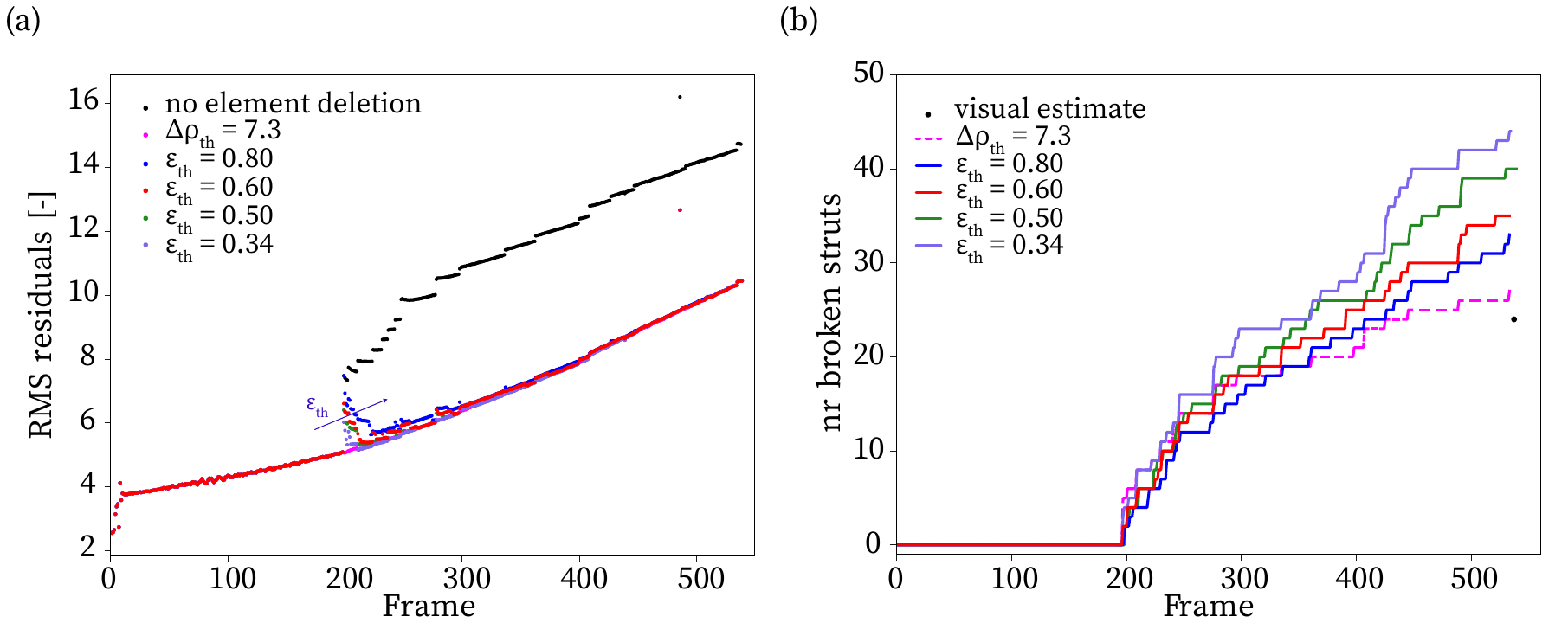}
    \caption{Analysis of the impact of the strain threshold $\epsilon_{th}$ on the global RMS residuals and on the number of damaged struts obtained from DIC with strain-based element deletion. The threshold $\epsilon_{th}$ is varied, and the DIC results compared with the reference obtained using residual-based element deletion. (a)~The residual-based criterion yields the most pronounced drop in global RMS residuals, followed by the lowest value of $\epsilon_{th}$ considered here. (b)~Residual-based element deletion results in fewer detected broken struts than any strain-based threshold; increasing $\epsilon_{th}$ reduces the number of detected damaged struts. The black dot corresponds to the 24 broken struts counted visually in the last image. The high number of broken struts observed for $\epsilon_{th}$ = 0.34 arises from the way struts are defined in the mesh, namely, each strut includes one-sixth of the adjacent node. Consequently, when a node fails, the associated struts are automatically counted as broken.}
    \label{fig:12}
\end{figure}

The global residuals drop sensitivity to the selected $\epsilon_{th}$ threshold is investigated by varying its value systematically around its estimates, and re-running the DIC analysis with integrated damage for each case. In our representative case, DIC with integrated element deletion is performed for different values of $\epsilon_\mathrm{th}$ ranging from 0.34 to 0.8, and both the reduction in global RMS residuals relative to the classical analysis with no element deletion, and the change of the number of broken struts over time are evaluated. A lower value ($\epsilon_\mathrm{th}$ = 0.2) led to a number of broken struts at the end of test above 100.
All tested values of $\epsilon_\mathrm{th}$ consistently detect the onset of damage at frame 200, in agreement with the results obtained with $\Delta\rho_{th}$. The lowest threshold $\epsilon_{th} = 0.34$ produces the most pronounced reduction in RMS residuals immediately after crack initiation, compared with the other tested values of $\epsilon_{th}$ (Figure~\ref{fig:12}(a)).

As expected, higher thresholds lead to weaker reductions in the global residuals, since fewer elements reach the critical strain during the test.
We note that the data-based element-removal method led to larger drops in the RMS global residuals than any of the strain thresholds considered herein (Figure~\ref{fig:12}(a)).

Regarding the number of broken struts, the data-driven method provides the closest agreement with the visual estimate. In contrast, all strain-based thresholds markedly overestimate the number of failed struts. This discrepancy arises from the less localized nature of damage detection when using a mechanics-based threshold, which derives from mesh continuity and from the post-processing procedure that assigns lattice-joint elements to the six adjacent struts. The choice of an appropriate strain threshold should therefore be guided by the objective of the analysis and may be informed when a robust value of $\epsilon_{th}$ is available from experiment, or is numerically estimated.

%
\bibliography{your_bibliography.bib}

\end{document}